\documentclass[a4paper,12pt]{article}

\usepackage{amsmath}
\usepackage{latexsym}

\newcommand{\cg}{C(G)}

\newcommand{\fdual}{\tilde{F}}

\newcommand{\first}{\ensuremath{\mbox{1}^{\mbox{\tiny st}}}}

\newcommand{\cc}{^*}

\newcommand{\gtd}{\itindex{\Gamma}{th.dyn.}}

\newcommand{\kahler}{\mathcal{K}}

\newcommand{\lomega}{\bra{\Omega}}

\newcommand{\lrcovsl}{{\,\stackrel{\leftrightarrow}{\!\!D \negthinspace
      \negthinspace \negthinspace \negthinspace /}}\mbox{\hspace{0.17ex}}}

\newcommand{\medsp}{\\[0.7ex]}

\newcommand{\nc}{\ensuremath{N_c} }
\newcommand{\nf}{\ensuremath{N_f} }

\newcommand{\migf}{\half{1-\gamma_5}}

\newcommand{\plgf}{\half{1+\gamma_5}}

\newcommand{\romega}{\ket{\Omega}}
\newcommand{\second}{\ensuremath{\mbox{2}^{\mbox{\tiny nd}}}}

\newcommand{\su}{\mbox{\slshape SU}}
\newcommand{\suppot}{\mathcal{W}}

\newcommand{\thev}{\theta_V}

\newcommand{\uone}{\mbox{\slshape U}(1)}

\newcommand{\Lindex}[1]{\ensuremath{\smallindex{\mathcal{L}}{#1}}}
\newcommand{\Ltext}[1]{\ensuremath{\itindex{\mathcal{L}}{#1}}}

\newcommand{\bra}[1]{\langle #1 |}
\newcommand{\dega}{\ensuremath{^\dag}}

\newcommand{\diff}[1]{\mbox{d}#1}

\newcommand{\half}[1]{\ensuremath{\frac{#1}{2}}}
\newcommand{\intd}[1]{\int \!\! #1 \;}
\newcommand{\inv}[1]{\ensuremath{\frac{1}{#1}}}
\newcommand{\ket}[1]{| #1 \rangle}
\newcommand{\metr}[1][]{g_{\varphi \bar{\varphi} #1}}

\newcommand{\pathd}[1]{\mathcal{D} #1 \;}

\newcommand{\wphi}[1][]{\suppot_{,\varphi #1}}
\newcommand{\bwphi}[1][]{\bar{\suppot}_{,\bar{\varphi} #1}}

\newcommand{\derfrac}[2][]{\frac{\partial #1}{\partial #2}}

\newcommand{\itindex}[2]{\ensuremath{#1_{\mbox{\scriptsize{\itshape #2}}}}} 
\newcommand{\smallindex}[2]{\ensuremath{#1_{\scriptscriptstyle{#2}}}}

\newcommand{\varfrac}[2][]{\frac{\delta #1}{\delta #2}}

\newcommand{\bddn}[3][ ]{\ensuremath{\Bar{#2}_{{#1} \dot{#3}}}}

\newcommand{\nddn}[3][ ]{\ensuremath{#2_{{#1} #3}}}

\DeclareMathOperator{\re}{Re}
\DeclareMathOperator{\im}{Im}
\DeclareMathOperator{\tr}{Tr}
\DeclareMathOperator{\hc}{h.c.}

\begin{document}
\bibliographystyle{phaip}
%
%
%
\thispagestyle{empty}
\begin{titlepage}

\vspace*{-1cm}
\hfill \parbox{3.5cm}{BUTP-2000/05} \\
\vspace*{1.0cm}

\begin{center}
  {\Large {\bf \hspace*{-0.2cm}Dynamical Symmetry Breaking
      \protect\vspace*{0.2cm} \\
      \protect\hspace*{0.2cm} and Static Limits of
      \protect\vspace*{0.2cm} \\      
      \protect\hspace*{0.2cm} Extended Super-Yang-Mills Theories}\large\footnote{Work
      supported in part by the Schweizerischer Nationalfonds.}\Large:
      \protect\vspace*{0.2cm} \\
      \protect\hspace*{0.2cm} \large{\bf A non-Seiberg-Wittian Approach}}
  \vspace*{2.5cm} \\

{\bf
    L. Bergamin and P. Minkowski} \\
    Institute for Theoretical Physics \\
    University of Bern \\
    CH - 3012 Bern, Switzerland
   \vspace*{0.8cm} \\  

March 13, 2000

\vspace*{3.0cm}

\begin{abstract}
\noindent
From a supersymmetry covariant source extension of $N=2$ SYM we study
non-trivial thermodynamical limits thereof. Using an argument by one of us
about the solution of the strong CP problem and the uniqueness of the QCD ground state we find that the dependence of
the effective potential on the defining field operators is severely
restricted. In contrast to the solution by Seiberg and Witten an acceptable
infrared behavior only exists for broken supersymmetry while the gauge
symmetry remains unbroken.
\end{abstract}
\end{center}

\end{titlepage}
%
%
\section{Introduction}
In the last few years numerous new results considering supersymmetry in a
field theoretical or string background have been derived. Although
supersymmetry and superstrings are the theoretical favorites for new physics
and for a consistent quantization of gravity, phenomenologically interesting models
have a serious problem: at the time present no supersymmetry breaking
mechanism in a field theoretical context is known.

As perturbative breaking mechanisms are excluded and since non-perturbative
regions are not available for exact calculations the situation is mostly
unclear. Several arguments have been given that rigid supersymmetry does not break
non-perturbatively. Witten \cite{witten82} argued that supersymmetry does not
break in certain classes of interesting models. Veneziano and Yankielowicz
\cite{veneziano82} concluded for pure $N=1$ Yang-Mills theory that supersymmetry remains unbroken after the
breakdown of chiral symmetry and Shifman and Vainshtein \cite{shifman88}
calculated the gluino condensate of $\su(N)$ theories exactly. Considering extended supersymmetry Seiberg and
Witten used the duality argument to derive an exact Wilsonian low-energy
effective action of $N=2$ SYM with and without matter
\cite{seiberg94,seiberg94:2}. They concluded that supersymmetry remains
unbroken while the gauge group is broken leading to non-trivial monopole
configurations.

In \cite{bib:mamink,bib:markus} it has been argued that the conclusion of
Veneziano and Yankielowicz might be wrong and that chiral symmetry breaking
induces supersymmetry breaking. The key leading to this different result is an
old observation by one of us \cite{minkowski78:1,minkowski78:2} that the
uniqueness of the non-perturbative ground state can solve the strong CP
problem setting the vacuum angle to zero. As a complete argument thereof never
has been published we discuss this topic in the Appendix of this paper. After
this modification the Witten index calculation breaks down and the
Veneziano-Yankielowicz effective action gets modified in such a way that supersymmetry breaks
together with chiral symmetry.

In this paper we want to extend the work of \cite{bib:mamink,bib:markus} to
gauge theories with two supersymmetries. Without going into the details of
$N=1$ we also want to clarify some points that have been omitted in
\cite{bib:mamink,bib:markus}.  The aim is to calculate a thermodynamical limit of our theory leading to
relations among vacuum expectation values of different composite
operators. Three steps lead to this result: After a short review of classical
and perturbative aspects of supersymmetry we define an external field
expansion of our system (section \ref{sec:minsource}). This expansion necessarily breaks the symmetries of
the theory (especially supersymmetry) but is done in a supersymmetry
covariant way. In section \ref{sec:action} we define an effective action in terms of the operators
associated to the external sources. Finally we relax the external fields in
the thermodynamical limit and we obtain different consistency conditions
among possible spontaneous parameters (section \ref{sec:breaking}). The main result is similar to $N=1$:
Unbroken supersymmetry does not allow for any condensates that can be attached
to its Lagrangian in a supersymmetry covariant way. This relates in $N=1$ the
gluino condensate, in $N=2$ all scalar condensates
to supersymmetry breaking. In contrast to \cite{seiberg94} we conclude
that supersymmetry is broken while the gauge symmetry remains unbroken.
\section{Basics about Pure $\boldsymbol{N=2}$ Yang-Mills Theory}
We briefly want to review some basic facts about extended supersymmetric
Yang-Mills theories. The anti-commutators among the charges of extended
supersymmetry are given by
\begin{align}
   \{ \nddn{Q^i}{\alpha} , \bddn[j]{Q}{\alpha} \} &= \delta^i_j P_{\alpha \dot{\alpha}} &  \{
    \nddn{Q^i}{\alpha} , \nddn{Q^j}{\beta} \} &= \varepsilon_{\alpha \beta} Z^{[ij]} & \{ \bddn[i]{Q}{\alpha} ,
    \bddn[j]{Q}{\beta} \} &= \varepsilon_{\dot{\alpha} \dot{\beta}} \bar{Z}_{[ij]}
\end{align}
To get theories with unbroken gauge-symmetry at tree-level the defining
algebra must have vanishing central charges. Then the algebra has an internal $\uone \otimes \su(2)$ symmetry that we represent according to
\begin{align}
  [Q_\alpha^i, R ]  &=  Q_\alpha^i & [Q_\alpha^i, I_r] &= \half{{(\tau_r)^i}_j}
  Q_\alpha^j & [I_r , I_s ] &= i \epsilon_{rst}  I_t
\end{align}
In superspace (central basis) charges and covariant derivatives can be represented as
\begin{subequations}
\begin{align}
  Q_\alpha^i &= i \partial_\alpha^i - \half{1} \sigma^\mu_{\alpha
  \dot{\alpha}} \bar{\theta}^{i \dot{\alpha}} \partial_\mu & \bar{Q}_{i
  \dot{\alpha}} &= - i \bar{\partial}_{i \dot{\alpha}} + \half{1}
  \theta^\alpha_i \sigma^\mu_{\alpha
  \dot{\alpha}} \partial_\mu \medsp
  D_\alpha^i &= \partial_\alpha^i - \half{i} \sigma^\mu_{\alpha
  \dot{\alpha}} \bar{\theta}^{i \dot{\alpha}} \partial_\mu & \bar{D}_{i
  \dot{\alpha}} &= -  \bar{\partial}_{i \dot{\alpha}} + \half{i}
  \theta^\alpha_i \sigma^\mu_{\alpha
  \dot{\alpha}} \partial_\mu
\end{align}
\end{subequations}
The canonical variables are then $ z^M = (x^\mu, \theta^{\alpha}_i,
\bar{\theta}^i_{\dot{\alpha}})$ with the following conjugation and
multiplication rules
\begin{subequations}
\begin{align}
  (\theta_{i \alpha})\cc &= \bar{\theta}^i_{\dot{\alpha}} &
  (\theta_{\alpha}^i)\cc &= - \bar{\theta}_{i \dot{\alpha}} && \medsp
  \theta^{\alpha}_i \theta^{\beta}_j &=  - \varepsilon_{ij} \theta^{\alpha \beta}
  - \varepsilon^{\alpha \beta} \theta_{ij} & \theta^4 &= \inv{12} \theta^\alpha_i \theta^{i \beta}
\theta_{j \alpha} \theta_\beta^j & \vartheta_\alpha^i &= \partial_\alpha^i \theta^4
\end{align}
\end{subequations}
where we use the spinorial metric $ \varepsilon_{\alpha \beta} =
\varepsilon^{ij} = \begin{pmatrix} 0 & 1 \\ -1 & 0 \end{pmatrix}$.

Supersymmetric field strengths and the classical invariant action can be
written in terms of a gauge chiral, scalar superfield $W$ that is subject to the constraint
$[\nabla^{ij} , W] = [\bar{\nabla}^{ij} , \bar{W} ]$ \cite{grimm78}, where
$\nabla^{ij}$ is the gauge covariant and symmetric version of the quadratic covariant
derivative. The gauge transformations of $W$ and its conjugate are
\begin{align}
  W\dega &= e^{-i X} \bar{W} e^{iX}&W &\rightarrow e^{i \Lambda} W e^{-i \Lambda} & W\dega &\rightarrow e^{i
  \bar{\Lambda}} W\dega e^{-i \bar{\Lambda}} & e^{i X} &\rightarrow e^{i
  \Lambda} e^{i X} e^{-i \bar{\Lambda}}
\end{align}
As in $N=1$ the component expansion is best written in a special gauge similar
to WZ gauge. In this gauge $W$ is not only gauge-chiral but also
chiral. Moreover the purely $\theta$ or $\bar{\theta}$
dependent terms of $X$ can be gauged away \cite{grimm78}. The expansion then
reads
\begin{equation}
  W(x, \theta^\alpha_i) = \sqrt{2} C(x) + \sqrt{2} \theta^\alpha_i \lambda_\alpha^i(x) +
  \theta^{\alpha \beta} v_{\alpha \beta}(x) + \theta_{ij} H^{ij}(x) +
  \vartheta^\alpha_i \chi_\alpha^i(x) + \theta^4 D(x)
\end{equation}
with the two highest components
\begin{align}
  \chi^\alpha_i &= \sqrt{2} i \bar{\sigma}_\mu^{\dot{\alpha} \alpha} [ D^\mu ,
  \bar{\lambda}_{i \dot{\alpha}} ] + i [ \bar{C} , \lambda^\alpha_i
  ] &  D &=  \sqrt{2} \bigl[ D^\mu , [D_\mu , \bar{C} ] \bigr] - \inv{\sqrt{2}} \bigl[ \bar{C} ,
  [\bar{C} , C ] \bigr] - i \{ \bar{\lambda}^i_{\dot{\alpha}} ,
  \bar{\lambda}_i^{\dot{\alpha}} \}
\end{align}
To define a proper action we introduce the complex coupling constant $\tau =
\inv{g^2} + \frac{i \vartheta}{8 \pi^2}$ and set  $S =
- \inv{8 \cg} \intd{\diff{^4x}}(\tau  \tr(W^2) + \hc) = - \inv{4 \cg}
\intd{\diff{^4x}}\re[\tau  \tr(W^2)]$ with the Lie-algebra invariant $\tr t^a
t^b = \cg \delta^{ab}$. The complete on-shell Lagrangian then reads \cite{grimm78}
\begin{equation}
\label{eq:invlag}
  \begin{split}
    \mathcal{L} &= \inv{\cg} \tr\bigl( \frac{1}{g^2} [ D_\mu , \bar{C}] [D^\mu , C] +
  \frac{i}{g^2}  \lambda^{i \alpha} \sigma^\mu_{\alpha
  \dot{\alpha}} [D_\mu, \bar{\lambda}_i^{\dot{\alpha}}] -  \inv{4 g^2} F_{\mu \nu} F^{\mu \nu} -
  \frac{\vartheta}{32\pi^2}  F_{\mu \nu} \tilde{F}^{\mu \nu}\medsp
  &\quad + \inv{4 g^2} H_{\{ij\}} H^{\{ij\}} +  \frac{1}{g^2} C [C, \bar{C} ] \bar{C} +  \frac{i}{\sqrt{2} g^2} C \{ \bar{\lambda}^i_{\dot{\alpha}} ,
  \bar{\lambda}_i^{\dot{\alpha}} \} -  \frac{i}{\sqrt{2} g^2} \bar{C} \{ \lambda_i^\alpha ,
  \lambda^i_\alpha \} \bigr)
  \end{split}
\end{equation}

To quantize the theory we have to express it in terms of unconstrained
superfields. This has been done by Howe et al.\ \cite{howe84} using central
basis. A simpler but still manifestly $N=2$ invariant formulation has been
given by Galperin et al.\ \cite{galperin84,galperin85:1,galperin85:2} with the
concept of harmonic superspace. According to the non-renormalization theorem
of extended supersymmetry \cite{grisaru82} Howe et al.\ derived from the
ghost-structure that $N=2$ SYM is perturbatively finite above one loop. Using
background fields in harmonic superspace \cite{buchbinder97} a more rigorous
proof of this statement has been given in \cite{buchbinder97:2}. Due to the
existence of $N=2$ SYM to any order in perturbation theory the defining
superfield $W$ has a definite meaning in quantum theory when replacing the
classical fields by properly renormalized ones. This will allow us to write
the effective potential in terms of the latter. Though we do not indicate this
explicitly we will assume in the following all fields to be renormalized.

The symmetries of a supersymmetric theory can be expressed in superspace by
means of the supercurrent. Extending the formulation of \cite{clark78} to
$N=2$ the current conservation including a chiral anomaly field can be written
as
\begin{equation}
\label{eq:twocurcon}
  {w^i}_j \Gamma = - \frac{i}{4} (\bar{\sigma}^\mu)^{\dot{\alpha} \alpha}
  \partial_\mu [D^i_\alpha , \bar{D}_{j \dot{\alpha}}] V - \delta^i_j (D^4 S -
  \bar{D}^4 \bar{S})
\end{equation}
where ${W^i}_j = \intd{\diff{^4 x}} {w^i}_j$ is the covariant operator
superfield of $N=2$ supersymmetry
\begin{align}
  {W^i}_j &= - \half{1} {\delta^i}_j R + {(\tau_r)^i}_j I_r - i \theta^{i
  \alpha} Q_{j \alpha} + i \bar{\theta}_{j \dot{\alpha}} \bar{Q}^{i
  \dot{\alpha}} + \theta^{i \alpha} P_{\alpha \dot{\alpha}} \bar{\theta}_j^{\dot{\alpha}}
\end{align}
and $V$ and $S$ are supercurrent and anomaly introduced by Sohnius
\cite{sohnius79}. Using the component structure thereof \cite{fisher83}
\begin{equation}
  \begin{split}
    V &= \theta^\alpha_i \bigl( - \half{1} \delta^i_j \sigma^\mu_{\alpha
    \dot{\alpha}} R_\mu + {(\tau_r)^i}_j \sigma^\mu_{\alpha
    \dot{\alpha}} (I_r)_\mu \bigr) \bar{\theta}^{j \dot{\alpha}} + (\theta_i
    \sigma^\mu \bar{\theta}^i)\bigl( (\theta_j \chi_\mu^j) + (\bar{\theta}^j
    \bar{\chi}_{\mu j})\bigr)\medsp
    &\quad + \half{1} (\theta_i \sigma^\mu \bar{\theta}^i)
    (\theta_j \sigma^\nu \bar{\theta}^j) v_{\mu \nu} + \ldots \medsp
    D^4 S - \bar{D}^4 \bar{S} &= F - \bar{F} - i (\bar{\theta}_i \bar{\sigma}
    \partial_\mu \psi^i) - i (\theta^i \sigma^\mu \partial_\mu \bar{\psi}_i) +
    \half{i} (\theta_i \sigma^\mu \bar{\theta}^i) \partial_\mu (F + \bar{F}) +
    \ldots
  \end{split}
\end{equation}
current conservations and anomalies are given by
\begin{subequations}
\begin{align}
     \partial^\mu R_\mu &= i w^{\mbox{\tiny R}} \Gamma + 4 \im F &
     \partial^\mu (I_r)_\mu &= i w_r^{\mbox{\tiny I}} \Gamma && \medsp
     v_{\mu \nu} &= - T_{\mu \nu} + \half{1} g_{\mu \nu} {T^\lambda}_\lambda &
     \partial^\nu T_{\mu \nu} &= i w_\mu^{\mbox{\tiny P}} \Gamma  & {T^\lambda}_\lambda
  &= -2 \re F \medsp
  Q_{\mu i} &= -i \chi_{\mu i} + \half{i} \sigma_\mu \bar{\sigma}_\nu \chi^{\nu}_i
  & \partial^\mu Q_{\mu i} &= i w_i^{\mbox{\tiny Q}} \Gamma & \bar{\sigma}^\mu Q_{\mu i}
  &= -2 i \bar{\psi}_i
\end{align}
\end{subequations}

In pure supersymmetric Yang-Mills theory the supercurrent is $V = - \inv{\cg}
\tr W \bar{W}$ while the anomaly is proportional to the action superfield. The
singlet axial current anomaly of this system reads
\begin{equation}
  \inv{\cg} \partial_\mu \tr(\lambda^i \sigma^\mu \bar{\lambda}_i) = -
  \frac{\nc}{8 \pi^2} \inv{\cg} \tr( F \fdual)
\end{equation}
The trace anomaly on the other hand is given by ${T^\mu}_\mu = \frac{\beta}{2
  g^3} \inv{\cg} \tr(F^2)$. The $\beta$-function of $N=2$ SYM is given to all
  orders in perturbation theory by \cite{jones75,howe83,novikov83} $\beta = - \frac{\nc}{8 \pi^2
  g^3}$. Noting that $R_\mu = - J_\mu^5 + \ldots$ we then get $F =
  \frac{\nc}{32 \pi^2 \cg}\tr (F_{\mu \nu} F^{\mu \nu} + i F_{\mu \nu}
  \fdual^{\mu \nu})$ and thus

  \begin{equation}
    S = \frac{\nc}{32 \pi^2 \cg} \tr W^2 = - \frac{\beta}{4 g^3 \cg} \tr W^2
  \end{equation}
Chiral, trace and supertrace anomaly become:
\begin{equation}
\label{eq:anomalies}
  \begin{split}
    \partial^\mu R_\mu &= \frac{\nc}{8 \pi^2 \cg} \tr \Bigl( (F_{\mu \nu}
  \fdual^{\mu \nu}) - 2 i \partial^\mu \bigl(C [D_\mu, \bar{C}] - [D_\mu , C]
  \bar{C}\bigr) - 2 D^\mu (\lambda^i \sigma_\mu \bar{\lambda}_i)\Bigr) \medsp
  {T^\mu}_\mu &= \frac{\beta}{g} \mathcal{L} \medsp
  \sigma^\mu Q_{\mu i} &= \frac{\sqrt{2} \beta}{g^3 \cg} \tr\bigl( \half{1}
  \lambda_i \sigma^{\mu \nu} F_{\mu \nu} - i \chi_i
  C - i \lambda^j H_{ij} \bigr)
  \end{split}
\end{equation}
For a detailed discussion of the component structure of the $N=2$ SYM current
see e.g. \cite{fisher84}.


\section{The Minimal Source Extension of the $\boldsymbol{N=2}$-System\label{sec:minsource}}
In this section we want to discuss how to introduce supersymmetry covariant sources to
the $N=2$ SYM-Lagrangian to be able to study thermodynamical limits of
composite operators. As unbroken supersymmetry exists in a
finite volume only with trivial boundary conditions, these sources break the
SUSY invariance of the Lagrangian as the highest component of a superfield. We
can nevertheless introduce SUSY covariant sources by replacing the complex
coupling constant $\tau$ by a complete superfield. In $N=1$ SYM
this has been discussed in detail in \cite{bib:mamink,bib:markus}. Considering
the thermodynamical limit we choose the sources non-vanishing in a volume $\itindex{V}{sub} \subset V$
and take the limit $\itindex{V}{sub} \subset V \rightarrow \infty$. We call
a source-term global if it is non-vanishing and constant inside
$\itindex{V}{sub}$ during this limiting process. For a detailed discussion of
this and other possible limits see \cite{bib:mamink,bib:markus}.

We start with the invariant Lagrangian in $N=2$ superspace \eqref{eq:invlag}:
\begin{align}
        \Lindex{0} &=  \intd{\diff{^4 \theta}} \tau \Phi + \hc & \tau &=
        \inv{g^2} + \frac{i \vartheta}{8 \pi^2} & \Phi &= - \inv{8 \cg} \tr W^2 & \tr t^a t^b &= \cg \delta^{ab}
\end{align}
Of course we have to add a gauge-fixing Lagrangian when considering the theory
as a quantum theory. As we do not add external sources to any of the
operators appearing therein we suppress this part of the action. The invariant Lagrangian can also be written as an integral over
full superspace which is important when considering the renormalization
property thereof. The above form however allows us to introduce a chiral source
multiplet instead of a full one. We will see that this minimal version is
enough to get sources of all relevant composite operators.

We thus replace the coupling constant $\tau$ by a chiral $N=2$ multiplet $J$
and the covariant source Lagrangian is then given by $\Ltext{J} =
\intd{\diff{^4 \theta}} J \Phi + \hc$. The full superfield $\Phi$ reads
\begin{equation}
\label{eq:sourcemultiplet}
\begin{split}
        \Phi &= - \inv{\cg} \tr\bigl[ \inv{4} C^2 + \half{1} \theta_i^\alpha \lambda_\alpha^i C + \inv{4} \theta_{ij}(\sqrt{2} H^{ij} C - \lambda^{i \alpha} \lambda^j_\alpha) +\inv{4} \theta^{\alpha \beta} (\frac{i}{\sqrt{2}} \sigma^{\mu \nu}_{\alpha \beta} F_{\mu \nu} C - \lambda_{i \alpha} \lambda^i_\beta) \medsp
        &\quad + \inv{2 \sqrt{2}} \vartheta^\alpha_i (\chi^i_\alpha C +
        \frac{i}{2} \lambda^{i \beta}\sigma^{\mu \nu}_{\alpha \beta} F_{\mu
        \nu} - \lambda_{j \alpha} H^{ij}) - \half{1} \theta^4 L \bigr]
\end{split}
\end{equation}
and we write for the chiral source multiplet:
\begin{equation}
        J(x) = \tau(x) + \theta^\alpha_i \zeta_\alpha^i - 4 \theta_{ij} m^{ij}(x) + \theta^{\alpha \beta} w_{\alpha \beta} + \vartheta^\alpha_i \kappa_\alpha^i + 4 \theta^4 M^2(x) 
\end{equation}
The non-scalar sources are needed to keep SUSY covariance. As these sources
break Poincar\'{e} invariance, their thermodynamical limit must be trivial and
we thus suppress them in the following. The highest component of our minimal source-Lagrangian then becomes
\begin{equation}
        \begin{split}
                \Ltext{J} &= \inv{\cg} \tr\bigl[ \re(\tau L) - M^2 C^2 -
                \bar{M}^2 \bar{C}^2 - m_{ij} (\sqrt{2} H^{ij} C - \lambda^{i \alpha} \lambda^j_\alpha) - \bar{m}_{ij} (\sqrt{2} H^{ij} \bar{C} - \bar{\lambda}^i_{\dot{\alpha}} \bar{\lambda}^{j \dot{\alpha}}) \bigr] 
        \end{split}
\end{equation}
The first term is the source of the quantum mechanical Lagrangian
\eqref{eq:invlag}. The complete source-extended Lagrangian is now given as $\Ltext{tot} = \Lindex{0} + \Ltext{J} + \Ltext{GF}$.

The source multiplet is subject to the constraint $J(x) \rightarrow 0\ (x
\rightarrow \infty)$. SUSY covariance enforces to take the limit of all
components in $J$ simultaneously while the relative normalizations thereof can
be changed by an appropriate supersymmetry transformation. Considering only
the highest components of all relevant superfields we can however modify this
picture and include $\Lindex{0}$ in $\Ltext{J}$ by changing the boundary conditions to
\begin{align}
        \lim_{x \rightarrow \infty} \tau(x) &= \tau & \lim_{x \rightarrow \infty} j(x) &= 0 &&\text{for all other components of $J(x)$}
\end{align}

In presence of a non-trivial source the auxiliary field $H_{ij}$ does not vanish, but the full auxiliary-field Lagrangian reads
\begin{align}
\label{eq:tripbasis}
        \Ltext{aux} &= - \inv{8 g^2} H^a_A H^a_A + \inv{\sqrt{2}}m_A H^a_A C^a + \inv{\sqrt{2}}\bar{m}_A H^a_A \bar{C}^a & {H^i}_j &= \half{{(\tau_A)^i}_j} H_A^a t^a
\end{align}
In this basis $H_A$ is anti-hermitian $(H_A)\dega = - H_A$ and $\bar{m}_A = -
m_A\dega$. Eliminating the auxiliary-fields we get
\begin{equation}
        \Ltext{aux} = - \frac{2 g^2}{\cg} \tr\bigl[ (m^{ij} C + \bar{m}^{ij} \bar{C}) (m_{ij} C + \bar{m}_{ij} \bar{C}) \bigr]
\end{equation}

Starting with a non-trivial source-configuration we can in principle obtain any
other configuration by applying a suitable SUSY transformation. This is not
problematic when considering local sources only. Arbitrary global sources
however can lead to unstable configurations. To avoid this problem we have to
introduce the following constraints on the lowest and highest component of $J$
($\mu^2 = - m_A m_A$, $\rho^2 = - m_A \Bar{m}_A$):
\begin{align}
        \re(\tau) &\geq 0 &  g^2 \rho^2 &\geq |M^2 + g^2 \mu^2|
\end{align}
We get the second constraint by noting that the eigenvalues of the scalar mass
matrix are given by $m_{1,2} = 2 \bigl(g^2 \rho^2 \pm |M^2 + g^2 \mu^2|\bigr)$ and that the ``$\Phi^4$'' term $\tr\bigl( C
[C,\bar{C}] \bar{C} \bigr)$ cannot stabilize negative mass terms.

Due to these constraints we can no longer apply any finite supertranslation to our Lagrangian. But as long as the constraints are inequalities we can still apply an arbitrary finite translation with small enough parameters.


\section{The Static Effective Action and its Symmetries\label{sec:action}}
In order to be able to perform the thermodynamical limit of the the source-extended
system we must formulate the
effective action in terms of the composite operators needed. The
SUSY covariance of this effective action can then be used to derive relations
between the thermodynamical limits of different operators. Formally this is
done by a Legendre transformation of the energy-functional:
\begin{equation}
\label{eq:funcionals}
  \begin{split}
    Z[J,\bar{J}] &= \intd{\pathd{X}} \exp(i S_0 + i S_J) = \exp\bigl( i W[J,\bar{J}] \bigr)
    \medsp
    \Gamma[\tilde{J},\Tilde{\Bar{J}}] &= \intd{\diff{^4 x}}\bigl( J(x) \varfrac[W{[}J{]}]{J(x)} +
    \hc \bigr) - W[J]
  \end{split}
\end{equation}
The variations and thermodynamical limiting conditions are
\begin{align}
  \varfrac{\tilde{J}(x)} \Gamma[\tilde{J},\Tilde{\Bar{J}}] &= J(x) \rightarrow 0
  & \varfrac{J(x)} W[J,\bar{J}] &= \tilde{J}(x) \rightarrow \tilde{J}\cc(x)
\end{align}
where a non-zero component of $\tilde{J}\cc(x)$ indicates the appearance of a
spontaneous parameter (vacuum expectation value).

When transferring the coupling constant $\tau$ to the boundary conditions we
may define instead of \eqref{eq:funcionals}
\begin{equation}
\label{eq:functionals2}
\begin{split}
  \Gamma[\tilde{J},\Tilde{\Bar{J}}] &= \intd{\diff{^4 x}}\bigl( (J(x) + \tau) \varfrac[W{[}J{]}]{J(x)} +
    \hc \bigr) - W[J] \medsp
  \varfrac{\tilde{J}(x)} \Gamma[\tilde{J},\Tilde{\Bar{J}}] &= J'(x) \rightarrow \tau
\end{split}
\end{equation}
The internal symmetries of the supersymmetry algebra define two Ward-Identities, one of
them being anomalous:
\begin{align}
   W_r^{\mbox{\tiny I}}(x) \Gamma &= 0 &  W^5(x) \Gamma &\sim F\fdual \cdot \Gamma &
   W^{\mbox{\tiny R}}(x) \Gamma &\sim \bigr(F\fdual + (\mbox{tot. der.})\bigl) \cdot \Gamma
\end{align}
As explained in the Appendix the anomalous chiral symmetry gets
restored when evaluated with respect to the ground-state, which sets the
(local and global) variations of $W[J,\bar{J}]$ with respect to the vacuum-angle to zero:
\begin{align}
\label{eq:gtdsym}
  W^5(x) \gtd &= W^{\mbox{\tiny R}}(x) \gtd = 0 &
  \varfrac[W{[}J,\bar{J}{]}]{\vartheta(x)} = \derfrac[W{[}J,\bar{J}{]}]{\vartheta} &= 0
\end{align}
This restored symmetry then implies that $\im \tau\cc(x) \sim \im \lomega L
\romega = \im \itindex{L}{cl} = 0$. 

\subsection{The Effective Potential as Static Part of a Nonlinear $\sigma$-Model\label{sec:sigmamodel}}
Besides the formal definition given above we can get $\gtd$ or the effective
potential by extracting the static part from the most general Lagrangian
obeying the symmetries of our theory, i.e. the most general Lagrangian of the
chiral $N=2$ multiplet of equation \eqref{eq:sourcemultiplet}. We extend this problem and derive
the Lagrangian of $k$ chiral multiplets.
\subsubsection{The most general Lagrangian of $k$ chiral $N=2$ superfields}
The nonlinear $\sigma$-models of $k$ chiral $N=1$ multiplets with or without
an additional second supersymmetry are well known to lead to hyper-K\"{a}hler and
K\"{a}hler manifolds, respectively \cite{zumino79,alvarez81}. The situation we are dealing with
here is slightly different. This can easily be seen when reducing a scalar
chiral $N=2$ superfield to the $N=1$ superfield formulation:
\begin{equation}
  \Phi = \Sigma_1(x, \theta_1) + \theta_2^\alpha
  \phi_\alpha (x, \theta_1) + \theta_2^2 \Sigma_2 (x, \theta_1)
\end{equation}
We can look at the scalar chiral $N=2$ superfield as a chiral $N=1$ superfield
depending itself on chiral $N=1$ superfields. As chiral superfields are irreducible
representations of supersymmetry we cannot split up the spinor field into
two (or more) scalar superfields. In a $N=1$ superspace formulation our
nonlinear $\sigma$-model thus necessarily involves spinorial superfields.

In the following it will however be easier to use directly the $N=2$
superspace formulation. The most general Lagrangian is of the form
\begin{align}
  \mathcal{L} &= \intd{\diff{^8 \theta}} \kahler(\Phi^A, \bar{\Phi}^{\bar{A}}) +
  \bigl( \intd{\diff{^4 \theta}} \suppot(\Phi^A) + \hc \bigr) &A&= 1, \ldots ,
  k
\end{align}
Using the component expansion
\begin{equation}
  \Phi = \varphi + \theta^\alpha_i \Lambda^i_\alpha + \theta^{\alpha \beta}
  v_{\alpha \beta} + \theta_{ij} C^{ij} + \vartheta^\alpha_i \eta^i_\alpha +
  \theta^4 D
\end{equation}
we can write the nonlinear $\sigma$-model (after integrating out superspace) in the following short form
\begin{equation}
\label{eq:genkahler}
  \begin{split}
    \kahler(\Phi, \bar{\Phi})|_{\theta^8} &= \Bigl[ D^A \partial_A +
    \Phi_0^{AB}  \partial_{AB} + \Phi_0^{ABC} \partial_{ABC} +
    \Phi_0^{ABCD}
    \partial_{ABCD} \Bigr]\Bigl[\hc\Bigr] \kahler\bigl(\varphi, \bar{\varphi}\bigr) \medsp
    &\quad + i \Bigl[ (\eta^{i \alpha})^A \partial_A +
    (\Phi_1^{AB})^\alpha \partial_{AB} +
    (\Phi_1^{ABC})^\alpha \partial_{ABC} \Bigr] \sigma^\mu_{\alpha \dot{\alpha}} \partial_\mu
    \medsp
    &\quad\quad\Bigl[ (\bar{\eta}^{\dot{\alpha}}_i)^{\bar{A}} \partial_{\bar{A}} +
    (\bar{\Phi}_1^{\bar{A}\bar{B}})^{\dot{\alpha}} \partial_{\bar{A}\bar{B}} +
    (\bar{\Phi}_1^{\bar{A}\bar{B}\bar{C}})^{\dot{\alpha}} \partial_{\bar{A}\bar{B}\bar{C}} \Bigr] \kahler\bigl(\varphi(x|),\bar{\varphi}(x)\bigr) \medsp
        &\quad - \Bigr[ (C^{ij})^A \partial_A + (\Phi_{1.1}^{AB})^{ij} \partial_{AB} \Bigr] \Box \Bigr[ (\bar{C}_{ij})^{\bar{A}} \partial_{\bar{A}} + (\bar{\Phi}_{1.1}^{\bar{A}\bar{B}})_{ij} \partial_{\bar{A}\bar{B}} \Bigr] \kahler\bigl(\varphi(x|),\bar{\varphi}(x)\bigr) \medsp
        &\quad - \Bigr[ (v^{\alpha \beta})^A \partial_A + (\Phi_{1.2}^{AB})^{\alpha \beta} \partial_{AB} \Bigr] \sigma^\mu_{\alpha \dot{\alpha}} \sigma^\nu_{\beta \dot{\beta}} \partial_\mu \partial_\nu \medsp
                &\quad \quad\Bigr[ (\bar{v}^{\dot{\alpha} \dot{\beta}})^{\bar{A}} \partial_{\bar{A}} + (\bar{\Phi}_{1.2}^{\bar{A}\bar{B}})^{\dot{\alpha} \dot{\beta}} \partial_{\bar{A}\bar{B}} \Bigr] \kahler\bigl(\varphi(x|),\bar{\varphi}(x)\bigr) \medsp
        &\quad -i (\Lambda^{i \alpha})^A \partial_A \; \sigma^\mu_{\alpha \dot{\alpha}} \partial_\mu \Box \; (\bar{\lambda}^{\dot{\alpha}}_i)^{\bar{A}} \partial_{\bar{A}} \kahler\bigl(\varphi(x|),\bar{\varphi}(x)\bigr) \medsp
                &\quad + \varphi^A \partial_A (\Box \Box \bar{\varphi}^{\bar{A}}) \partial_{\bar{A}} \kahler\bigl(\varphi(x),\bar{\varphi}(x)\bigr) 
  \end{split}
\end{equation}
with the components
\begin{align*}
  \Phi_0^{AB} &= \half{1} (v^{\alpha \beta})^A  (v_{\alpha
    \beta})^B - \half{1} (C_{ij})^A (C^{ij})^B - (\eta^\alpha_i)^A
    (\Lambda_\alpha^i)^B &&\medsp
    \Phi_0^{ABC} &= \half{1} \bigl(
    (C_{ij})^A (\Lambda^{i \alpha})^B (\Lambda^{j}_\alpha)^C - (v^{\alpha
    \beta})^A (\Lambda_{i \alpha})^B (\Lambda^{i}_\beta)^C\bigr) &
    \Phi_0^{ABCD} &= \inv{12} (\Lambda^\alpha_i \Lambda^{i
    \beta} \Lambda_{j \alpha} \Lambda^j_\beta )^{ABCD}\medsp
        (\Phi_1^{AB})^\alpha &= - \bigl( (v^{\alpha \beta})^A (\Lambda^i_\beta)^B + (\Lambda^\alpha_j)^A (C^{ij})^B\bigr) & (\Phi_1^{ABC})^\alpha &= \inv{3} (\Lambda^\alpha_j \Lambda^{j \beta} \Lambda^i_\beta)^{ABC} \medsp
        (\Phi_{1.1}^{AB})^{ij} &= - \half{1} (\Lambda^{i \alpha})^A (\Lambda^j_\alpha)^B & (\Phi_{1.2}^{AB})^{\alpha \beta} &= - \half{1} (\Lambda_i^\alpha)^A (\Lambda^{i \beta})^B
\end{align*}
$\partial_{A_1 \ldots A_n}$ stands for $\frac{\partial^n}{\partial \varphi^{A_1} \ldots \partial \varphi^{A_n}}$ and the vertical line in $\kahler\bigl(\varphi(x|),\bar{\varphi}(x)\bigr)$ indicates that the space-time derivatives only act on the $\bar{\varphi}$-fields inside $\kahler(\varphi, \bar{ \varphi})$.

In a completely analogous way we get the expression for the highest component of the superpotential:
\begin{equation}
\label{eq:genpotential}
        \suppot(\Phi)|_{\theta^4} = \Bigl[ D^A \partial_A +
    \Phi_0^{AB} \partial_{AB}  + \Phi_0^{ABC} \partial_{ABC} +
    \Phi_0^{ABCD} \partial_{ABCD}
    \Bigr] \suppot(\varphi)
\end{equation}
Although the complete Lagrangian of this model is rather complicated when expressed on component level we can immediately read off some important properties:
\begin{itemize}
        \item The terms quadratic in the fields $D$, $\eta^i_\alpha$, $C^{ij}$ and $v_{\alpha \beta}$ are given by
        \begin{equation}
                \begin{split}
                \Lindex{2} = \frac{\partial^2 \kahler}{\partial \varphi^A \partial \bar{\varphi}^{\bar{A}}} \bigl( &D^A \bar{D}^{\bar{A}} + i (\eta^{i \alpha})^A \sigma^\mu_{\alpha \dot{\alpha}} \partial_\mu (\bar{\eta}^{\dot{\alpha}}_i)^{\bar{A}} - (C^{ij})^A \Box (\bar{C}_{ij})^{\bar{A}}\medsp
                & - (v^{\alpha \beta})^A \sigma^\mu_{\alpha \dot{\alpha}} \sigma^\nu_{\beta \dot{\beta}} \partial_\mu \partial_\nu (\bar{v}^{\dot{\alpha} \dot{\beta}})^{\bar{A}} \bigr) + \ldots
                \end{split}
        \end{equation}
        To get stable $p^2$-fluctuations of the dynamical fields in the above equation, $\kahler$ must be K\"{a}hlerian and $g_{A \bar{A}} = \frac{\partial^2 \kahler}{\partial \varphi^A \partial \bar{\varphi}^{\bar{A}}}$ defines the hermitian metric of the manifold. In contrast to the $N=2$ matter fields the manifold need not be hyper-K\"{a}hlerian. This is in fact easy to understand: Both $N=2$ matter fields, the Howe-Stelle-Townsend as well as the Fayet-Sohnius hypermultiplet, are in a non-trivial representation of the internal $\su(2)$ symmetry. This induces the quaternionic structure of a hyper-K\"{a}hler manifold when constructing a nonlinear $\sigma$-model. Our superfield transforms with respect to the trivial representation and consequently such a structure is not needed.
        \item $\Lindex{2}$ does not generate correct $p^2$-terms for the
          remaining fields $\varphi$ and $\Lambda^i_\alpha$. This can already
          be seen from dimensional considerations and leads to important constraints on the dynamics of the system, as such a term must be produced from higher order components. We can indeed read off from the $\partial_{AB\bar{A}\bar{B}} \kahler$ component the expression:
          \begin{equation}
            \begin{split}
              \Lindex{4} = \frac{\partial^4 \kahler}{\partial \varphi^A \partial
              \varphi^B \partial \bar{\varphi}^{\bar{A}} \partial
              \bar{\varphi}^{\bar{B}}} \bigl(& i (C_{ji})^B
              (\bar{C}^{ik})^{\bar{B}} (\Lambda^j)^A \sigma^\mu \partial_\mu
              (\bar{\Lambda}_k)^{\bar{A}}\medsp
              &- (C_{ji})^B
              (\bar{C}^{ij})^{\bar{B}} \varphi^A \Box \bar{\varphi}^{\bar{A}}
              \bigr) + \ldots
            \end{split}
          \end{equation}
          This expression defines a second metric of our system which does not
          transform trivially under the internal $\su(2)$. It is given by:
          \begin{equation}
            {(\tilde{g}_{A \bar{A}})_j}^k = (C_{ji})^B g_{A \bar{A}, B \bar{B}} (\bar{C}^{ik})^{\bar{B}}
          \end{equation}
          It immediately follows that the $p^2$-fluctuations of this system
          are only stable if at least one operator $(C_{ji})^B
          (\bar{C}^{ik})^{\bar{B}}$ has a non-trivial vacuum-expectation
          value. The hermiticity of this metric follows trivially. Notice that
          from equation \eqref{eq:genkahler} a similar
          expression  $\sim v_{\alpha \beta} \bar{v}^{\dot{\alpha}
          \dot{\beta}}$ containing the correct number of derivatives can be
          obtained. But due to its Lorentz structure it cannot contribute to
          this metric.

          The most significant consequence of this second metric and of the
          associated vacuum-expectation value is not its pure existence, but
          the fact that it breaks the internal
          $\su(2)$ symmetry of the theory. We will discuss this point again
          when specializing to our SYM-model.
        \item We want to eliminate the auxiliary fields. From
          \eqref{eq:genkahler} and \eqref{eq:genpotential} we can read off the
          following expression for the variation with respect to $\bar{D}^{\bar{A}}$:
          \begin{equation}
            \label{eq:secgenaux}
            g_{A \bar{A}} D^A = - g_{A \bar{A},B} \Phi_0^{AB}  - g_{A \bar{A},
    BC} \Phi_0^{ABC} - g_{A \bar{A}, BCD} \Phi_0^{ABCD} - \derfrac[\bar{\suppot}]{\bar{\varphi}^{\bar{A}}}
          \end{equation}
\end{itemize}

Now we specialize to a single chiral superfield and extract the static
part of the Lagrangian given above. Suppressing again all contributions from
non Lorentz-scalars we get
\begin{equation}
  \begin{split}
    \Ltext{th.dyn.} &=  D \metr \bar{D} - \half{1} D \bar{C}_{kl} \bar{C}^{kl} \metr[,
    \bar{\varphi}]  - \half{1}  C_{ij} C^{ij} D \metr[, \varphi] + \inv{4} C_{ij} C^{ij} \bar{C}_{kl} \bar{C}^{kl} \metr[,\varphi
    \bar{\varphi}] \medsp
    &\quad + D \wphi + \bar{D} \bwphi - \half{1} C_{ij} C^{ij} \wphi[\varphi] -
    \half{1} \bar{C}_{kl} \bar{C}^{kl} \bwphi[\bar{\varphi}]
  \end{split}
\end{equation}
The effective potential we are looking for is (up to a sign) the above
expression when identifying the components with the classical fields $\lomega
\Phi \romega$, where $\Phi$ has been defined in equation
\eqref{eq:sourcemultiplet}.

Finally we want to define the geometrical objects of our K\"{a}hler
manifold. As the dual metric of a one-dimensional manifold is simply given by $g^{\varphi \bar{\varphi}} =
\inv{\metr}$, connection and curvature are:
\begin{align}
  \Gamma^{\varphi}_{\varphi \varphi} &= \Gamma = \frac{\metr[,\varphi]}{\metr}
  & \bar{\Gamma}^{\bar{\varphi}}_{\bar{\varphi} \bar{\varphi}} &= \bar{\Gamma}
  = \frac{\metr[,\bar{\varphi}]}{\metr} & R_{\varphi \bar{\varphi} \varphi
  \bar{\varphi}} &= R = \metr[,\varphi \bar{\varphi}] - \frac{\metr[,\varphi] \metr[,\bar{\varphi}]}{\metr}
\end{align}


\section{SUSY (Non-)Breaking Conditions from the Static Effective
Action\label{sec:breaking}}
Finally we combine the results of the previous sections. Using the component
structure of the effective potential and
the symmetries of the latter from its formal definition we get constraints on
possible spontaneous parameters. It is well known that the order parameter of
supersymmetry breaking is the vacuum energy density due to the relation
\begin{equation}
\label{eq:susybreak}
  \half{1} \lomega \bigl( (\sigma_\mu)_{\alpha \dot{\alpha}} \{ Q^{i \alpha} ,
  \bar{S}^{\dot{\alpha}}_{j \nu} \} + \bar{\sigma}_\mu^{\dot{\alpha} \alpha}
  \{\bar{Q}_{j \dot{\alpha}}, S^i_{\nu \alpha} \bigr) \romega = \delta^i_j \lomega T_{\mu \nu}
  \romega = E_0 \delta^i_j g_{\mu \nu}
\end{equation}
which is directly connected to $\lomega \mathcal{L} \romega$ via
equation \eqref{eq:anomalies}. As $\itindex{L}{cl}$ is one of our defining
variables, we can easily control possible SUSY-breaking
effects. According to equation \eqref{eq:susybreak} it seems to be impossible to break rigid supersymmetry partially. But in a
spontaneously broken theory we can consider the algebra of currents only. The latter can be
modified such that partial supersymmetry breaking is indeed possible in some models
\cite{hughes86,hughes86:2,bagger94,bagger97,antoniadis96}. However we will not
discuss this here.
\subsection{The Chiral-Weight Puzzle}
As the chiral (or R-) symmetry is unbroken in classical SYM, every classical
superfield of this theory has a well-defined chiral weight. In quantum theory
however chiral symmetry is broken. In the static limit it gets
restored at least for the Lagrangian multiplet and it should thus re-inherit
its classical chiral weight\footnote{Some points in the subsequent discussion
  have been omitted in \cite{bib:mamink,bib:markus}. Especially the chiral weight of the
  superpotential has been left open and the possibility of a non-trivial
  dependence on the imaginary part of the coupling constant has not been discussed. All arguments
  given here also hold --mutatis mutandis-- in $N=1$.}.

When assigning chiral weight $+1$ to the left-handed super-generator
$\chi[Q_\alpha^i] = +1$, we get the classical weights: $\chi[W] = -2$,
$\chi[\Phi] = -4$, $\chi[J] = 0$. The first two weights follow from the fact
that the classical Lagrangian has vanishing chiral weight, the last from
$\chi[\tau] = 0$. Restoration of the chiral symmetry now tells us that
$\chi[\itindex{\Phi}{cl}] = -4$, especially $\chi[\itindex{L}{cl}] = 0$ which
follows from equation \eqref{eq:gtdsym} that can alternatively be written
as $\im \itindex{L}{cl} \equiv 0$. More complicated is the behavior of the
source-multiplet under quantization. Considering the coupling constant $\tau$
we can define:
\begin{align}
  \itindex{\tau}{QM} &= \inv{g^2} + \frac{i \vartheta}{8 \pi^2} &
  \itindex{\tau}{V} &= - \frac{i \itindex{\vartheta}{V}}{8 \pi^2} & \tau(x) &= \inv{g^2(x)} + \frac{i \vartheta(x)}{8 \pi^2}
\end{align}
The first $\tau$ is the usual renormalized coupling constant, the second one
the vacuum angle and the last one the source term. Due to the dynamical equations
\eqref{eq:genlim} and \eqref{eq:genlim2} in the Appendix the effective
quantum-mechanical coupling constant $\itindex{\tau}{eff} = \itindex{\tau}{QM}
+ \itindex{\tau}{V}$ transforms under chiral rotations as:
\begin{align}
  Q &\rightarrow e^{i \alpha} Q & \itindex{\tau}{eff} &\rightarrow
  \itindex{\tau}{eff} + \frac{2 i \alpha}{8 \pi^2}
\end{align}
Thus $\itindex{\tau}{eff}$ has a logarithmic weight:
$\chi\bigl[\exp(\itindex{\tau}{eff})\bigr] = \inv{4 \pi^2}$. After
transforming the effective coupling constant to the boundary conditions the
lowest component of the source-multiplet therefore has a special weight.

It is now easy to read off the chiral weights of the static K\"{a}hler- and
superpotential. Noting that $\chi[\varphi] = -4$ and thus $\derfrac{\varphi} =
+4$ we get
\begin{align}
  \chi[\gtd] &= 0 & \chi[\kahler] &= 0 & \chi[\suppot] &= - 4
\end{align}

The fact that $\gtd$ has a defined chiral weight restricts the dependence
of the latter on the classical fields. Before going into details we however
want to eliminate the auxiliary field of the nonlinear $\sigma$-model.
\subsection{Elimination of the Second-Generation Auxiliary Fields}
Our model includes two types of auxiliary fields: $D \sim \itindex{L}{cl}$ and
the auxiliary field of the underlying theory $H^{ij}$. We
will refer to them as \second- and \first-generation auxiliary fields
respectively. The \second-generation auxiliary fields are eliminated using equation \eqref{eq:secgenaux}. Taking the static part we obtain
\begin{equation}
  \begin{split}
\label{eq:effpot}
    D &= \lomega \frac{g^2}{2} \mathcal{L} \romega = \half{1} \Gamma C_{ij}
    C^{ij} - \frac{\bwphi}{\metr} \medsp
    \gtd &= \frac{|\wphi|^2}{\metr} + \half{1} \bigl(C_{ij}C^{ij}
    (\wphi[\varphi] - \Gamma) + \hc \bigr) - \inv{4} C_{ij}C^{ij} \bar{C}_{kl}
    \bar{C}^{kl} R
  \end{split}
\end{equation}
After elimination the auxiliary field $\gtd$ can depend on the $\su(2)$ singlets
$C_{ij}C^{ij}$, $C_{ij}\bar{C}^{ij}$, $\varphi$ and its hermitian
conjugate only. Besides the combinations of these fields with chiral weight zero we
could also construct terms of the form $\exp(16 \pi^2
\itindex{\tau}{eff}) \varphi$. Such a term is however excluded by the
invariance of the theory under a global change of the $\vartheta$-angle, as
from \eqref{eq:gtdsym}
\begin{align}
   \derfrac[W{[}J,\bar{J}{]}]{\vartheta} &= 0 &\Rightarrow&& \derfrac[\Gamma{[}\itindex{\Phi}{cl}{]}]{\vartheta} &= 0
\end{align}
which may be checked explicitly by using the inverse of equation
\eqref{eq:funcionals} and noting that
\begin{equation}
\intd{\diff{^4x}} \bigl(\intd{\diff{^4 \theta}} \itindex{\Phi}{cl}
\derfrac{\vartheta} J(x) + \hc \bigr) = 0
\end{equation}
Also note that it is irrelevant which $\vartheta$
(quantum-mechanical, vacuum-angle or global source) we choose in the above
equation, as the latter only appear in the specific combination of the equations
\eqref{eq:genlim} and \eqref{eq:genlim2} respectively. We thus conclude that after eliminating
$D$ and after turning off all sources the static effective action can only
depend on
\begin{equation}
  \gtd[\itindex{\Phi}{cl}] = \gtd\Bigl[ |\varphi|^2, C_{ij}\bar{C}^{ij},
  C_{ij}C^{ij} \bar{C}_{kl}\bar{C}^{kl}, \bigl\{ \varphi
  \bar{C}_{kl}\bar{C}^{kl}, \frac{C_{ij}C^{ij}}{\varphi}, \mbox{and} \hc\bigr\}
  \Bigr]
\end{equation}

We have already seen that the fields $C_{ij}$ cannot acquire a
vacuum expectation value due to the internal $\su(2)$ symmetry. This
observation contradicts the conclusion made in section \ref{sec:sigmamodel}. We want to argue here that our model nevertheless exists
in the thermodynamical limit. The assumptions we are making are the existence
of the underlying theory ($N=2$ SYM) as non-perturbative field theory and its
stability with respect to small perturbations of all composite operators
considered here. Then the effective action in terms of these composite
operators (the components of $J(x)$) must indeed exist and is given by the limiting
process $J(x) \rightarrow 0$. If $J(x)$ is non-zero the non-equilibrium
effective action has the form of the discussed nonlinear $\sigma$-model due
its SUSY covariance. Together with the above assumption the limiting
process must be defined and leads to the correct effective action. What does
this mean for the dependence of $\gtd$ on $C_{ij}$? If $\varphi \neq 0$ in the
limit, $\gtd$ can only depend on positive powers of $C_{ij}C^{ij}$, if $\varphi
= 0$ a definite limit of $\frac{C_{ij}C^{ij}}{\varphi}$ could in principle
exist. Besides the fact that such a term immediately sets all scalar
condensates to zero to ensure the stability of the effective action under
variations (see next section), SUSY-transformations on the non-equilibrium
system can e.g.\ set $m_{ij}(x)$ to zero while $M^2(x) \neq 0$, which excludes
the dependence on such a fraction. Thus $\gtd$ can only depend on positive
powers of $C_{ij}C^{ij}$.
\subsection{Elimination of the First-Generation Auxiliary Fields}
The elimination of the \first-generation auxiliary fields is somewhat
different from the usual procedure, as they appear inside the expression of a
classical composite operator. This situation is similar to a
(non-supersymmetric) theory of two scalar fields, one of them being
auxiliary. The Lagrangian shall be given by:
\begin{align}
  \mathcal{L} &= \half{1} \partial_\mu \phi \partial^\mu \phi + \half{1} F^2 -
  V(\phi) & \lomega \phi^2
\romega|_{\mbox{\tiny pert. theory}} &= 0 & \lomega \phi^2
\romega|_{\mbox{\tiny non-pert.}} &\neq 0
\end{align}
By attaching a source $(F\phi)(x)$ we get under the basic assumption that the
non-perturbative vacuum does not change under the elimination of the auxiliary fields
\begin{align}
  \lomega (F\phi)(x) \romega|_{m(x)} &= m(x) \lomega (\phi^2)(x) \romega &&
  \longrightarrow 0\ (m(x) \rightarrow 0)
\end{align}
The variation of $\lomega (F\phi)(x) \romega|_{m(x)}$ however is non-vanishing
by the assumption made above. Thus we get the following variation of the
energy-functional
\begin{subequations}
\begin{align}
  \varfrac[W{[}m{]}]{m(x)}\Bigl|_{m(x) \rightarrow 0}\Bigr. &= \itindex{(F\phi)}{cl} = 0
  \medsp
  \varfrac[^2 W{[}m{]}]{m(x) \delta m(y)}\Bigl|_{m(x) \rightarrow 0}\Bigr. &=
  \varfrac{m(y)} \itindex{(F\phi)}{cl}\Bigl|_{m(x) \rightarrow 0}\Bigr. = \delta(x-y) \itindex{(\phi^2)}{cl}
\end{align}
\end{subequations}
On the other hand we may calculate the variation of
$\Gamma[\itindex{(F\phi)}{cl}]$ with respect to the ''wrong'' variable $m(x)$:
$\varfrac{m(x)} \Gamma[\itindex{(F\phi)}{cl}] = \itindex{(F\phi)}{cl}$. Thus
all variations of $W$ and $\Gamma$ with respect to $m(x)$ are equivalent and we
get the non-vanishing condensate as second variation of the effective action:
\begin{subequations}
\begin{align}
  \varfrac[^2 \Gamma{[}\itindex{(F\phi)}{cl}{]}]{m(x) \delta m(y)}\Bigl|_{m(x)
  \rightarrow 0}\Bigr. &= \delta(x-y) \itindex{(\phi^2)}{cl} \medsp
  \varfrac[^n
  W{[}m{]}]{m(x_1) \ldots \delta m(x_n)} &= \varfrac[^n
  \Gamma{[}\itindex{(F\phi)}{cl}{]}]{m(x_1) \ldots \delta m(x_n)}
\end{align}
\end{subequations}
Of course we could also extract the physical condensate when varying the
effective action with respect to its defining variable
$\itindex{(F\phi)}{cl}$. This relation reads:
\begin{equation}
  \varfrac[^2 \Gamma{[}\itindex{(F\phi)}{cl}{]}]{\itindex{(F\phi)}{cl}(x)
  \delta \itindex{(F\phi)}{cl}(y)} = - \frac{\delta(x-y)}{\itindex{(\phi^2)}{cl}(x)}
\end{equation}
Finally we want to note that all formal relations given above remain true when
introducing any new fields and attaching any new sources.

To translate this to our SUSY-model we first rewrite the latter according to
equation \eqref{eq:tripbasis}. Then source-extension and Legendre transform of
the $H^A$ dependent $C^A$-fields read:
\begin{align}
  \mathcal{L}_{m^A} &= -2 (m^A C^A + \bar{m}^A \bar{C}^A) &
  \Gamma[\itindex{C}{cl}^A] &= - 2 \intd{\diff{^4 x}} (m^A  \itindex{C}{cl}^A
  + \bar{m}^A \itindex{\bar{C}}{cl}^A) - W[m^A]
\end{align}
The variations of the field and the effective action thus become:
\begin{subequations}
  \begin{align}
    \varfrac[C^B(z)]{m^A(x)} &= -g^2 \delta^B_A \delta(x-z)  \itindex{(C^2)}{cl}
    &  \varfrac[C^B(z)]{\bar{m}^A(x)} &= -g^2 \delta^B_A \delta(x-z)
    \itindex{( C\bar{C})}{cl} \label{eq:Cvar1}\medsp
    \varfrac[\bar{C}^B(z)]{m^A(x)} &= -g^2 \delta^B_A \delta(x-z) \itindex{( C\bar{C})}{cl}
     &  \varfrac[\bar{C}^B(z)]{\bar{m}^A(x)} &= -g^2 \delta^B_A \delta(x-z)
    \itindex{( \bar{C}^2)}{cl}\label{eq:Cvar2} \medsp
    \varfrac[\Gamma{[}\itindex{C}{cl}^A{]}]{m^A (x)} &= -2 \itindex{C}{cl}^A
    (x) & \varfrac[\Gamma{[}\itindex{C}{cl}^A{]}]{\bar{m}^A (x)} &= -2
    \itindex{\bar{C}}{cl}^A \medsp
    \varfrac[^2 \Gamma{[}\itindex{C}{cl}^A{]}]{m^A (x) \delta m^B(z)} &= 2g^2
    \delta^{AB} \delta(x-z) \itindex{( C^2)}{cl}  & \varfrac[^2 \Gamma{[}\itindex{C}{cl}^A{]}]{\bar{m}^A (x) \delta \bar{m}^B(z)} &= 2g^2
    \delta^{AB} \delta(x-z) \itindex{( \bar{C}^2)}{cl} \label{eq:Gvar1} \medsp
    \varfrac[^2 \Gamma{[}\itindex{C}{cl}^A{]}]{m^A (x) \delta \bar{m}^B(z)} &= 2g^2
    \delta^{AB} \delta(x-z) \itindex{( C \bar{C})}{cl} &&\label{eq:Gvar2}
  \end{align}
\end{subequations}

Besides these formal relations which follow directly from the
Legendre transformation we want to calculate the same variations using our
explicit effective potential. To do this we have to distinguish more carefully
two different variations with respect to the source $m^A(x)$. On
one hand the effective action may depend on the source as a function of the
classical fields $m^A = m^A[\itindex{\Phi}{cl}(x)]$ even without eliminating
the \first-generation auxiliary fields, on the other hand the $C^A$ depend
explicitly on $m^A(x)$ after the elimination. Therefore we expand the
effective action to second order in the fields and sources $\psi^A_i = \{C^A,
\bar{C}^A, m^A, \bar{m}^A \}$. This variation of the effective action (and completely analogous of the
K\"{a}hler- and the super-potential) is given by
\begin{equation}
  \Gamma[C^A,m^A,\ldots] = \Gamma^0 + \half{1} \intd{\diff{^4x}\diff{^4y}}
  \Bigl.\varfrac[^2 \Gamma]{\psi^A_i(x) \delta \psi^B_j(y)}\Bigr| \psi^A_i(x) \psi^B_j(y)
\end{equation}
As $\Gamma$ (and all other functions involved) is a $\su(2)$-singlet it only
depends on the quadratic combinations $\zeta_i(x,y) = \{C^A(x) C^A(y), m^A(x) C^A(y), m^A(x) m^A(y),
  \ldots\}$. Using the vanishing thermodynamical limits of the source and of
$C^A$ we can then rewrite the second variation as
\begin{equation}
  \begin{split}
    \intd{\diff{^4x}\diff{^4y}} \Bigl.\varfrac[^2 \Gamma]{\psi^A_i \delta
    \psi^B_j}\Bigr| \psi^A_i \psi^B_j &=
    \intd{\diff{^4x}\diff{^4y}\diff{^4u}\diff{^4v}}
    \Bigl. \varfrac[\Gamma]{\zeta_k(u,v)} \varfrac[^2\zeta_k(u,v)]{\psi^A_i(x) \delta
    \psi^B_j(y)}\Bigr| \psi^A_i(x) \psi^B_j(y)\medsp
    &= \intd{\diff{^4x}\diff{^4y}} \delta^{AB} \Gamma^{ij}(x,y) \psi^C_i(x) \psi^C_j(y)
  \end{split}
\end{equation}
Using equations \eqref{eq:Cvar1} and \eqref{eq:Cvar2} the variations of our
effective potential \eqref{eq:effpot} can now easily be calculated.
\subsection{Non-Vanishing Condensates and Supersymmetry Breaking}
We are now ready to discuss the restrictions on unbroken supersymmetry
from the effective potential and its variations. First of all equation \eqref{eq:effpot} reads after dropping all
trivial terms
\begin{align}
    \gtd\bigl[|\varphi|^2\bigr] &= \frac{|\wphi|^2}{\metr} & D &= \lomega \frac{g^2}{2} \mathcal{L} \romega =  - \frac{\bwphi}{\metr}
\end{align}
\subsubsection{Unbroken Supersymmetry}
Completely analogous to the discussion of the $N=1$ case
\cite{bib:mamink,bib:markus}, the effective potential attains its minimum
along a circle in the complex plane of the lowest component of the classical
Lagrangian superfield. Again analyticity of the superpotential $\wphi$ implies
that unbroken supersymmetry ($\Ltext{cl} \sim \bwphi = 0$) can only exist
non-trivially at $\varphi = 0$. All other solutions are trivial the sense that
$\wphi \equiv 0$  $\forall \varphi$ which is unacceptable when perturbing the
system with a source $M^2(x)$ while keeping $C_{ij} = 0$.

Considering the variations of the effective action \eqref{eq:Gvar1} and \eqref{eq:Gvar2} the curvature-term does
only contribute to fourth and higher order variations and consequently we only have to
look at
\begin{equation}
\label{eq:varaction}
  \gtd = \frac{|\wphi|^2}{\metr} - \inv{4} \bigl( C^A C^A (\wphi[\varphi] -
  \Gamma) + \bar{C}^A \bar{C}^A (\bwphi[\bar{\varphi}] - \bar{\Gamma})\bigr)
\end{equation}
From the first term we have to take its quadratic expansion as
discussed above, the second is non-trivial when varying both $C^A$-fields only. The former will vanish as all terms of
the expansion are still $\sim \wphi$ or $\sim \bwphi$ which vanishes by
assumption of unbroken SUSY. The second term is $\sim (\wphi[\varphi] -
  \Gamma)$, which is the derivative of a function with chiral weight zero with
  respect to $\varphi$. As such a function cannot have a term linear in
  $\varphi$ and as all non-constant terms vanish in the thermodynamical limit,
  this derivative must vanish at $\varphi = 0$. Thus it immediately
  follows that $\itindex{(C^2)}{cl} = \itindex{(C \bar{C})}{cl} = 0$.

This leads to the main conclusion of this paper: Unbroken supersymmetry does
not allow for any non-trivial condensates that can be attached to its
Lagrangian in a SUSY covariant way.
\subsubsection{Broken Supersymmetry}
If supersymmetry is broken ($\wphi \neq 0$) the chiral weight restricts the
superpotential to be of the form $\suppot = a + b \varphi$ and thus $\wphi =
\mbox{constant}$, $\wphi[\varphi] \equiv 0$ independently of the value of
$\wphi$ at the minimum. Thus the minimum of the effective potential is
completely defined by the maximum of the K\"{a}hler metric. Our formalism
only tells us that this maximum must be on a circle in the complex
$\varphi$-plane, but we cannot decide whether $\varphi = 0$ is the
correct solution or not. In contrast to $N=1$  where the
lowest component of the Lagrangian-multiplet is the essentially non-zero
gaugino-condensate, such a restriction does not exist in $N=2$. Especially our formalism
does not exclude the possibility of $N=2$ SYM being described in the
thermodynamical limit by a simple manifold with a single maximum at the origin.

If $\varphi = 0$ the variations
\eqref{eq:Gvar1} and \eqref{eq:Gvar2} lead to constraint-equations on the
expansion-coefficients of $\suppot$ and $\kahler$, as $\frac{|\wphi|^2}{\metr}$
leads to non-trivial contributions only. Lack of a detailed
knowledge of the the dependence of these two functions on the source $m^A$ in
the vicinity of the thermodynamical limit a non-trivial condensate of $C
\bar{C}$ is possible but not required. Of course these
consistency-conditions become much more complicated when considering $\varphi
\neq 0$ and they do not allow for any conclusions at this level of
calculations.

Besides the condensates discussed here other non-trivial vacuum expectation
values are of course possible. On one hand we have all non-renormalizable
operators which we do not want to discuss. On the other hand there are
renormalizable operators that do not appear in the Lagrangian superfield. Of
main interest is the question of a possible Higgs phenomenon due to a non-vanishing
vacuum expectation value of the scalars, leading to magnetic monopoles. Under
the assumption of unbroken
supersymmetry this has been discussed by Seiberg and Witten
\cite{seiberg94}. In contrast to their calculation  the field-strength tensor gets a non-trivial
vacuum expectation value when supersymmetry is broken. Striebel \cite{striebel87} showed 
that magnetic monopole configurations cannot exist in a constant background field. We thus conclude that
supersymmetry breaks without touching any other symmetries. Consequently the
only Goldstone modes are the two Goldstone fermions from SUSY breaking.


\section{Summary and Conclusions}
Using the covariant source-extension non-trivial thermodynamical limits of
supersymmetric theories can be studied in a supersymmetry covariant way. For
pure ($N=1$ and $N=2$) SYM theories the effective potential can be
derived. Together with the uniqueness of the ground state of non-Abelian gauge
theories with respect to the variation of the vacuum angle this links in $N=2$ the
condensate of the Lagrangian to those of the scalars. An acceptable infrared
behavior is consistent with broken supersymmetry only. Supersymmetry
breaking then lifts the classical and perturbative vacuum degeneracy at a
fixed modulus and monopole configurations disappear. Consequently the gauge
symmetry remains unbroken.

Interesting questions are left open: The existence of massless Goldstone
fermions is restricted by phenomenological results. The coupling of the theory to
supergravity thus has to be studied. Moreover only two special models have
been considered yet. A general statement about supersymmetry breaking in a
wide class of interesting models is not yet possible. Of particularly interest
is the fate of perturbatively finite theories like $N=4$ SYM under
non-perturbative quantum corrections.


\appendix
\section*{APPENDIX\\QCD, Strong CP and Thermodynamical
Limits\label{sec:strongcp}}
\setcounter{section}{1}
In this Appendix we want to explain in detail our arguments that allow us to
set in a QCD-like theory without explicit CP-violation except for the
topological term any CP-violating phase to zero. The calculations are
basically old ideas by one of us \cite{minkowski78:1,minkowski78:2} (see also
\cite{minkowski90,minkowski90:2}). As none of these citations contains a
complete discussion of all arguments, we give a rather detailed Appendix
considering this problem here.
\subsection{Non-Trivial Topology and the Singlet Anomaly}
As all calculations in this Appendix are --up to some uninteresting
constants-- independent of the representation of the fermions we consider QCD
with \nf quark flavors only. The free (Minkowskian) QCD Lagrangian is given
by
\begin{align}
  \Lindex{0} &= - \inv{4 \cg g^2} \tr(F_{\mu \nu} F^{\mu \nu}) + \half{i}\,
  \bar{\psi}_i \lrcovsl \psi_i &&\text{$i = 1, \ldots, \nf$}
\end{align}
The most general (complex) mass-Lagrangian in QCD is given by
\begin{align}
  - \Ltext{m} &= \bar{\psi}_i M_{ik} \plgf \psi_k + \bar{\psi}_i \bar{M}_{ik}
  \migf \psi_k & (M_{ik})\dega &= \bar{M}_{ki} & \plgf \psi &= \begin{pmatrix}
  \psi_L \\ 0 \end{pmatrix}
\end{align}
Gauge-fixing is done using the standard BRST procedure. We do not
derive this in detail here.

Non-Abelian gauge theories have a non-trivial topology. A topological
invariant of the Euclidean $\su(N)$ gauge theory is given by (Pontrjagin
index, second Chern-number, Atiyah-Singer index theorem)
\begin{align}
  Q=-C_2=\mbox{index}D_+ &= \inv{16 \pi^2 \cg} \int_{S^4} \tr F^2 = - \inv{48
  \pi^2 \cg} \int_{S^3} \tr(h_{-+} d h_{-+})^3 = n
\end{align}
In the ``physical'' language this represents the instanton number and reads (Euclidean space)
\begin{align}
  Q &= - \inv{32 \pi^2 \cg} \intd{\diff{^4 x}} \tr(F_{\mu \nu} \fdual^{\mu \nu})
  &&\text{$Q \in \mathbf{Z}$}
\end{align}
The physical meaning of the instantons is usually studied in the temporal
gauge $A^0 = 0$. In the vacuum states\footnote{Our terminology allows a
  theory to have several vacua (e.g. $\ket{N}$) but one unique ground state
  $\romega$ only.} $F_{\mu \nu} = 0$ the connections are
then spatially pure gauge $A_i = e^{- \alpha(\vec{x})} \partial_i
e^{\alpha(\vec{x})}$, where $\alpha(\vec{x})$ are traceless, anti-hermitian $N
\times N$ matrices. In this formulation the Pontrjagin index is equivalent to
\begin{align}
  Q &= N_+ - N_- & N_\pm &= - \inv{48 \pi^2 \cg} \intd{\diff{^3x}} \varepsilon^{ijk}
  \tr\bigl( (e^\alpha \partial_i e^{- \alpha}) (e^\alpha \partial_j e^{-
  \alpha}) (e^\alpha \partial_k e^{- \alpha}) \bigl)\Bigr|_{t = \pm \infty}
\end{align}
Gauge transformations changing the topological sector of a given
field configuration are called large gauge transformations. As the operators
of the large gauge transformations $U(g^k) \ket{N} = \ket{N+k}$ and the
Hamiltonian may be simultaneously diagonalized, the eigenvectors of $U$ and
$H$ must be of the form \cite{callan76,jackiw76}
\begin{align}
        \ket{\theta} &= \sum_N e^{i N \theta} \ket{N} & U(g^k) \ket{N} &= e^{ikN} \ket{N}
\end{align}
Different values of $\theta$ represent different sectors of the theory in the
sense that $\bra{\theta'} B \ket{\theta} = 0$ $(\theta \neq \theta')$ for
any gauge-invariant operator $B$. Without proof we note that the above
structure of the Yang-Mills vacuum is also present in quantum theory. Of
course the exact realization of both states $\ket{\theta}$ and $\ket{N}$ is then
unknown but also unimportant here.

Finally we get for the Euclidean generating functional
\begin{align}
  \label{eq:EucZ1}
        \lim_{t \rightarrow \infty} \bra{\theta'} e^{-Ht} \ket{\theta} &= 2 \pi \delta(\theta - \theta') Z_\theta & Z_\theta &= \sum_Q e^{-i Q \theta} \intd{\pathd{X_Q}} \exp(- \itindex{S}{Euc})
\end{align}
Treating $\theta$ as a free parameter its effect is to add the term
\begin{align}
  \label{eq:thetaterms}
        \Ltext{Euc} &\rightarrow \Ltext{Euc} + \frac{i \theta}{32 \pi^2 \cg}
        \tr F_{\mu \nu} \fdual^{\mu \nu} & \Ltext{Mink} &\rightarrow
        \Ltext{Mink} - \frac{\theta}{32 \pi^2 \cg} \tr F_{\mu \nu} \fdual^{\mu \nu}
\end{align}
to the Lagrangian and again using the path integral measure over all instanton configurations.

In the chiral limit QCD is classically invariant under $\su(\nf)_R \times
    \su(\nf)_L \times U(1) \times U(1)_A$. However $U(1)_A$ is anomalous with
    the (Minkowskian) anomaly \cite{hooft76, hooft76:2, callan76,
    jackiw76, rajaraman82}
\begin{align}
        \partial_\mu J^\mu_5 &= \partial_\mu (\bar{\psi}_i \gamma^\mu \gamma_5
        \psi_i) = 2 \nf q  + i \bar{\psi} \tilde{M} \psi & \tilde{M} &= \gamma_5 (M+\bar{M}) + (M - \bar{M})
\end{align}
where $Q = \intd{\diff{^4 x}} q(x)$ is the Pontrjagin index. The anomaly
allows to rotate complex phases of the mass matrix away by means of $\theta \rightarrow \bar{\theta} =
\theta + \arg \det M$.

The anomalous term is a total divergence of the Chern-Simons form: $\tr F^2 = d Q_3$ with $Q_3 = \tr(AdA + \frac{2}{3} A^3)$. Thus we can construct a new conserved current
\begin{align}
        F^\mu_5 &= J^\mu_5 + 2 \nf K^\mu & K^\mu &= \inv{16 \pi^2 \cg} \varepsilon^{\mu \nu \rho \sigma} \tr(A_\nu \partial_\rho A_\sigma + \frac{2}{3} A_\nu A_\rho A_\sigma)
\end{align}
with $\partial_\mu F^\mu_5 = 0$ in the chiral limit. However the new conserved
charge $Q_5 = \intd{\diff{^3x}} F_5^0$ is not gauge invariant, but $U_5 =
\exp(-i \frac{\pi}{\nf} Q_5)$ generates the discrete symmetry corresponding to
a shift $\theta \rightarrow \theta + 2 \pi$.


\subsection{Thermodynamical Constraints on the $\theta$-Parameter}
We now want to study the generating functional \eqref{eq:EucZ1} more in
detail. First we fix the remaining gauge freedom in the standard way by
imposing the constraints $A_i|_{t = - \infty} = 0$, $e^{\alpha(\vec{x})}
\rightarrow 1$, $(\vec{x} \rightarrow \infty)$. Then the generating functional
$Z_{\thev}$ is the sum over the vacuum-to-vacuum transition amplitudes from $N_- = 0$ to an
arbitrary $N_+ = N$:
\begin{equation}
\begin{split}
  Z_{\thev} &= \sum_N \exp(i N \thev)\, \bra{N} \intd{\pathd{X}}
  \exp(- \itindex{S}{Euc}) \ket{0}\medsp 
  &= \sum_N \exp(i N (\thev - \bar{\theta}))\, \bra{N} \intd{\pathd{X}} \exp(- S_0)
  \ket{0} = \sum_N \bra{N} \intd{\pathd{X}} \exp(- S_{\thev}) \ket{0}
\end{split}
\end{equation}
In the above equation $\bra{N}$ and $\ket{0}$ indicate the index of the states
at $t = \pm \infty$ and $\bar{\theta}$ is the effective coupling constant
$\bar{\theta} = \theta + \arg \det M$. Without loss of generality we
will assume in the following that all phases of the mass matrix have been
rotated into the $\theta$ parameter and thus $\bar{\theta} =
\theta$. Then the action $S_0$ is the usual QCD action without
$\theta$-term and $S_{\thev}$ is given by $S_{\thev} = S_0 + \frac{i
  (\theta - \thev)}{32 \pi^2 \cg} \intd{\diff{^4x}} \tr F \fdual$.

The crucial point in our discussion is the interpretation of the two
parameters $\thev$ and $\theta$. The way we defined them,
$\theta$ is the coupling constant of a renormalizable and gauge invariant
operator as the quark masses or the Yang-Mills coupling constant. Of course
this coupling constant may be chosen arbitrarily but fixed. The
parameter $\thev$ on the other hand is a free phase of an off-diagonal
S-matrix element. A priori it is also arbitrary but the dynamics of the
system may determine its value uniquely for a given set of coupling
constants. In our opinion $\thev$ actually has to be dynamical: For a given
set of external parameters (i.e. coupling constants) a theory must have an
unique ground-state. In order to satisfy this uniqueness $\thev$  must either
be irrelevant or dynamical as it does not belong to the set
of external parameters. Thus the overall coupling constant $\theta - \thev$ of the CP-violating operator
$F\fdual$ may indeed be subject to dynamical
constraints. Calculating the thermodynamical limit of the associated operator
we want to show that this is the case and that the dynamical value of
the $\thev$-parameter in a theory without explicit CP-breaking (except for
the $F\fdual$ term) is $\theta - \thev = 0$.

The interpretation of the different $\thev$-vacua as being caused by
tunneling between topological vacua $\ket{N}$ is a gauge-dependent
interpretation restricted to the usage of temporal gauge
\cite{bernard77}. However the existence of a free phase $\thev$ is not, as
shown in \cite{rothe79}. The existence of this free parameter is the
important difference between the ``normal'' coupling constants ($M_{ik}$, $g$)
and the ``topological'' coupling constant $\theta$.

Including the possible sources the generating functional may be written as
\begin{align}
  Z[J] &= \intd{\pathd{X}} e^{- \itindex{S}{\tiny Euc}} & \itindex{S}{Euc} &=
  \itindex{S}{QCD} + \itindex{S}{GF} +  \itindex{S}{J}
\end{align}
where we include a term $\sim \tr F\fdual$ with an arbitrary ``coupling
constant'' $\theta - \thev$ in $\Ltext{QCD}$. Considering the
sources we are mainly interested in those for local composite operators:
\begin{equation}
  \Ltext{J} = - \inv{4 \cg} \tau \tr(F_{\mu \nu} F_{\mu \nu}) - \inv{4 \cg} \vartheta \tr(F_{\mu
  \nu} \fdual_{\mu \nu}) + \bar{\psi}_i \sigma^{ij} \psi_j + \ldots
\end{equation}
All sources are subject to the boundary conditions $\lim_{x \rightarrow
  \infty} J(x) = 0$. For the gauge boson and the quark field configurations we
  introduce the boundary conditions
\begin{equation}
\begin{split}
    2 \varfrac{\tau(x)} Z[J] &= \langle \inv{2 \cg} \tr(F_{\mu
    \nu} F_{\mu \nu}) e^{-S}\rangle = \nc B^2 Z[J]  \medsp
  2 \varfrac{\vartheta(x)} Z[J] &= \langle \inv{2 \cg} \tr(F_{\mu
    \nu} \fdual_{\mu \nu}) e^{-S}\rangle = \nc B \tilde{B} Z[J] \medsp
  \varfrac{\sigma^{ij} (x)} Z[J] &= - \langle
    \bar{\psi}_i \psi_j e^{-S} \rangle = M^S_{ij} Z[J] \medsp
    \ldots& 
\end{split}
\end{equation}
The parameters $B$ and $\tilde{B}$ are related by the inequality $|\tilde{B}|
\leq B$. In terms of the energy functional and of the effective action we get the following variations with respect to the
sources and associated operators respectively:
\begin{align}
\label{eq:eqlim}
  \varfrac{J(x)} W[J] &= - \tilde{J}(x) & \varfrac{\tilde{J}(x)} \Gamma[\tilde{J}] &= - J(x)
\end{align}
In our case the associated operators are given by
\begin{align}
  \tilde{\tau}(x) &=  \lomega \inv{4 \cg} \tr(F_{\mu \nu} F_{\mu \nu} )\romega &
  \tilde{\vartheta}(x) &= \lomega \inv{4 \cg} \tr(F_{\mu \nu} \fdual_{\mu \nu})
  \romega && \ldots
\end{align}
As discussed in detail in \cite{minkowski90} a non-vanishing vacuum expectation value of a given local operator for vanishing
associated source leads to a spontaneous parameter. We indicate such a
parameter by a star. The spontaneous parameter $B\cc$ associated with the gluon
condensate would then be
\begin{align}
  \tilde{\tau}(x) &=  \lomega \inv{4 \cg} \tr(F_{\mu \nu} F_{\mu \nu}) \romega &
  \varfrac{\tilde{\tau}} \Gamma &= \derfrac{B} \Gamma|_{B = B\cc} =0
\end{align}

Operators are in general subject to renormalization. Considering quantum
effects we thus assume the above operators to be renormalized. The spontaneous parameters are then
functions of the (renormalization group invariant) generalized coupling
constants: $B^* = B^*(\itindex{\Lambda}{QCD}, m_s, \theta)$, $\tilde{B}^* =
\tilde{B}^*(\itindex{\Lambda}{QCD}, m_s, \theta)$. Since the above operator
gives (by hypothesis) the ground-state of the theory, $B\cc$ and
$\tilde{B}\cc$ must be of the form
\begin{align}
  B\cc &= C(\theta) (\itindex{\Lambda}{QCD})^2 & \tilde{B}\cc &= \tilde{C}(\theta) (\itindex{\Lambda}{QCD})^2
\end{align}
From the integrated variation with respect to $\vartheta$
\begin{align}
  \intd[V]{\diff{^4x}} \tilde{\vartheta}(x) &= - \derfrac{\vartheta}
  W[J]
\end{align}
we get the thermodynamical limit determining the value of $\tilde{B}\cc$:
\begin{align}
\label{eq:thlimit}
  \half{1} \nc B\cc \tilde{B}\cc(\vartheta) &= - \lim_{V \rightarrow \infty}
  \inv{V} \derfrac{\vartheta} W[\vartheta, \ldots]\bigr |_{J \rightarrow 0}
\end{align}

The explicit calculation of this limit is now similar to a limiting
process of a spin chain in statistical mechanics: We consider a spin-chain in a magnetic field
$\vec{B}$ of an arbitrary but fixed direction. The question is whether the
angle between the spins and the magnetic field  may be chosen non-trivially in the limit of an infinitely large
chain and zero temperature leading to the possibility of CP-violating
configurations. Calculating the non-trivial
limit, the stabilization of this situation during the limiting process would
require an infinite amount of energy with respect to a \emph{dynamical}
variable. Consequently non-trivial phases relax leading to a non-equilibrium
state violating the (approximate) translation invariance during the limiting
process. This flipping of spins in the limit $T \rightarrow 0$ also takes
place when considering the ground-state of a finite chain. There is however an
important difference between the situations with $l= (\mbox{finite})$ and $l \rightarrow
\infty$ respectively. While the transverse susceptibility is well defined at
a finite length $\derfrac[^n]{\theta^n} Z$ leads out of the Hilbert space for
$l\rightarrow \infty$. Thus the translation-invariance is
getting restored at the trivial angle
and the CP-violating configurations do not exist in the limit,
although we may choose an arbitrary angle $\theta$ between the spin and the
$\vec{B}$-field direction at the beginning. 

In the case of QCD we must be careful to choose a well-defined limit which
forces to have non-vanishing values for both $B$ and $M^S$. To ensure this we
choose sources $\tilde{\tau} \neq 0$ and $\tilde{\sigma} \neq 0$ inside a
sub-volume $\itindex{V}{sub}$, but vanishing sources on the complement $V
\setminus \itindex{V}{sub}$ and then take the infinite volume limit
$\itindex{V}{sub} \subset V \rightarrow \infty$, as discussed in detail in
\cite{bib:mamink}. Of course we have to take the limit $J \rightarrow 0$ for
all sources in the end, as the equilibrium conditions demand both equations in
\eqref{eq:eqlim} to be valid simultaneously. Then the spontaneous parameter $(M^S)\cc$ and $B\cc$
may vanish again. We do not want to discuss this purely dynamical problem here, as it does not
change our conclusions.

Considering additional infrared problems due to infinite correlation lengths we
just note that $L$ and $L'$ defined as
\begin{align}
  \lomega F(x) \mathcal{O}_8 F(y) \romega &\sim \exp( -
  \frac{z^2}{L^2}) &  \lomega F\fdual (x)  F\fdual(y) \romega &\sim \exp( -
  m_{\eta'} L')
\end{align}
with $z = x-y$, are both finite.

The leading terms of the energy-density and of $\tilde{\tau}$ are given by
\cite{hooft76:2,callan76} (dropping a possible constant independent of $\theta$)
\begin{align}
  \frac{E(\theta)}{V} &= - 2K e^{- S} \cos \theta & \tilde{\tau} &= - 64
  \pi^2 i K e^{-S} \sin \theta
\end{align}
where $K$ is a constant independent of $\theta$ and non-zero due to our choice
of sources. Replacing $\theta$ again by $\theta - \thev$ the absolute minima
are at
\begin{equation}
\label{eq:minthe}
  \thev = \theta \mod 2 \pi
\end{equation}
and in the thermodynamical limit the
free parameter takes one of these values removing all spontaneous CP
violation and thus setting $\tilde{B}\cc$ to zero. Thus we get the following system of
equilibrium conditions (still under assumption of a real mass term)
\begin{align}
  \varfrac[\Gamma]{\tilde{J}(x)} &= 0 & \theta - \thev &= 0 &
  \varfrac[W]{\vartheta(x)} &= 0 & \varfrac[W]{\tau(x)} &= - \half{1} \nc (B\cc)^2
\end{align}

If we allow again for a general mass matrix the effective action can only
depend on the combination $(\theta -\thev) + \arg \det M$. The dynamical
equations for $\thev$ are thus:
\begin{align}
\label{eq:genlim}
  \varfrac[W]{\vartheta(x)} &= 0 & (\theta -\thev) + \arg \det M &= 0
\end{align}
The dynamical constraints now restore the chiral invariance when evaluated
with respect to the ground-state and we get
\begin{align}
\label{eq:chiinv1}
  \lomega \partial_\mu J_5^\mu \romega &= 0 = i (M + \bar{M}) \lomega
  \bar{\psi} \gamma_5 \psi \romega + i (M - \bar{M}) \lomega \bar{\psi} \psi \romega
\end{align}
which is a nontrivial dynamical equation for the quark-condensates. By defining the operators $\tilde{\sigma}_L$ and $\tilde{\sigma}_R$ as
$\tilde{\sigma}_{L/R} = \bar{\psi} \half{1 \pm \gamma_5} \psi$ the above constraint becomes
\begin{align}
\label{eq:chiinv2}
  M \varfrac[W]{\sigma_L} &= \bar{M} \varfrac[W]{\sigma_R}
\end{align}
It is again obvious that the dynamics minimize CP violation, i.e.\ it minimizes the
relative angle between the mass-matrix and the quark-condensates. Moreover
equations \eqref{eq:genlim} and \eqref{eq:chiinv1} or \eqref{eq:chiinv2}
directly connect the phase of the quark-condensates to the effective
$\theta$-parameter appearing in the Lagrangian. To be explicit equation
\eqref{eq:chiinv2} tells us that $\arg \det M = \arg \det \tilde{\sigma}_R = -
\arg \det \tilde{\sigma}_L$ and thus we may rewrite equation \eqref{eq:genlim}
as
\begin{align}
\label{eq:genlim2}
  \varfrac[W]{\vartheta(x)} &= 0 & (\theta -\thev) + \arg \det \tilde{\sigma}_R &= 0
\end{align}
If we are considering a theory with massless quarks, we can still attach
non-trivial sources $\sigma_{L/R}$ and thus the thermodynamical
limit still connects the $\theta$-angle to the phase of the quark-condensate
as given in the equation above. As all calculations in this section are independent of the representation of
the quarks, the results are also applicable to supersymmetric gauge
theories.

Of course the question arises whether the partition function derived here
$Z(\theta \equiv 0) = Z_0$ is related to the partition function of the
standard interpretation restricted to the value $\theta = 0$
$Z(\theta)|_{\theta = 0}$. The discussion of this point is now completely
analogous to the above example of a spin chain. In the standard interpretation of global topological objects
$Z(\theta)$ is defined for all values of $\theta$ with a unique spectrum (in
the case of massless fermions it is even independent of $\theta$). Thus we may
safely travel along the circle of the $\theta$-phase and consequently all
variations $\frac{\delta^n}{\delta \theta^n} Z(\theta)$ are globally and
locally well defined (of
course we
assume here physically relevant changes of the $\theta$ angle and not just
variable transformations in the anomaly term).

The situation is however completely different in our calculation: At non-trivial values of $\theta$ the
path integral does not converge and thus the partition function need not even be
defined. Therefore global changes of the $\theta$-parameter lead out of the
Hilbert-space. This just means that a global change of the coupling-constant
$\theta$ leads to a dynamical reaction of $\thev$ such that the effective
parameter remains zero. Formally this can be written as a constraint on the
global variation with respect to $\theta$ (or $\thev$), namely
$\derfrac{\theta}W \equiv 0$ -- together with \eqref{eq:thlimit} this is just
another way to see that $\tilde{B}\cc$ must be zero.

This completely different mathematical behavior makes it reasonable that the
two limiting processes $\lim_{V \rightarrow \infty}$ and $\lim_{\theta
  \rightarrow 0}$ need not be interchangeable and thus in general $
Z_0 \neq Z(\theta)|_{\theta = 0}$.


\subsection{Concluding Remarks}
We have discussed in this Appendix how non-perturbative dynamics lead to a natural
solution of the strong CP problem without re-introducing the $U(1)$
problem. Considering this point we want to make two remarks. Witten
\cite{witten79} and Veneziano \cite{veneziano79} argued that there exists a relation between the mass of the
$\eta'$ and the topological susceptibility $\frac{\partial^2}{\partial \theta^2} W |_{\theta = 0}$. In
principle such a relation does not stand in contradiction to our analysis, however the
direct meaning of $\derfrac[^2]{\theta^2} W$ is far from clear, as we are leaving
the Hilbert space when going over to non-vanishing $\theta$'s. Notwithstanding
the mass-square of the $\eta'$ can be obtained through local variations $\varfrac{\vartheta}$.

It has also been proven by Shifman, Vainshtein and Zakharov \cite{shifman80}
that, if it were possible to start from a QCD Lagrangian with a non-trivial
$\theta$-term, dynamics can not resolve both the $U(1)$ and the strong CP
problem. However our procedure removes all $\theta$-angles ab initio through a
complete analysis of the thermodynamical limit showing that this assumption is not valid.

Our last remark considers a different suggestion to solve the strong CP
problem. It has been shown by Banerjee, Mitra and Chatterjee \cite{banerjee94}
that complex phases in the mass term may be decoupled from the $\theta$-term
by using a representation of Euclidean fermions different from the usual
Osterwalder-Schrader scenario. Although this apparently resolves the fine-tuning problem
in the Standard Model we are left with the unsatisfactory situation that there
would exist two fundamentally different version of QCD. Our analysis shows that this
need not be the case because in this alternative version of QCD non-trivial
$\theta$-angles relax thermodynamically, too. Though the two versions may be
different technically, they are equivalent after studying non-perturbative
dynamics.

\bibliography{biblio}

\begin{thebibliography}{10}

\bibitem{witten82}
E.~Witten,
\newblock Nucl. Phys. {\bf B202}, 253 (1982).

\bibitem{veneziano82}
G.~Veneziano and S.~Yankielowicz,
\newblock Phys. Lett. {\bf B113}, 231 (1982).

\bibitem{shifman88}
M.~A. Shifman and A.~I. Vainshtein,
\newblock Nucl. Phys. {\bf B296}, 445 (1988).

\bibitem{seiberg94}
N.~Seiberg and E.~Witten,
\newblock Nucl. Phys. {\bf B426}, 19 (1994).

\bibitem{seiberg94:2}
N.~Seiberg and E.~Witten,
\newblock Nucl. Phys. {\bf B431}, 484 (1994).

\bibitem{bib:mamink}
M.~Leibundgut and P.~Minkowski,
\newblock Nucl. Phys. {\bf B531}, 95 (1998).

\bibitem{bib:markus}
M.~Leibundgut,
\newblock Spontaneous {S}upersymmetry {B}reaking in {P}ure {S}uper
  {Y}ang-{M}ills {T}heories,
\newblock PhD thesis at the {U}niversity of {B}ern, 1998.

\bibitem{minkowski78:1}
P.~Minkowski,
\newblock The mass of the $\eta'(958)$, {CP} and {C} conservation in {QCD},
  manifestations of a {J}ospehson effect,
\newblock MPI-PAE/Pth 37/78.

\bibitem{minkowski78:2}
P.~Minkowski,
\newblock Phys. Lett. {\bf B76}, 439 (1978).

\bibitem{grimm78}
R.~Grimm, M.~Sohnius, and J.~Wess,
\newblock Nucl. Phys. {\bf B133}, 275 (1978).

\bibitem{howe84}
P.~S. Howe, K.~S. Stelle, and P.~K. Townsend,
\newblock Nucl. Phys. {\bf B236}, 125 (1984).

\bibitem{galperin84}
A.~Galperin, E.~Ivanov, S.~Kalitsyn, V.~Ogievetsky, and E.~Sokatchev,
\newblock Class. Quant. Grav. {\bf 1}, 469 (1984).

\bibitem{galperin85:1}
A.~Galperin, E.~A. Ivanov, V.~Ogievetsky, and E.~Sokatchev,
\newblock Class. Quant. Grav. {\bf 2}, 601 (1985).

\bibitem{galperin85:2}
A.~Galperin, E.~Ivanov, V.~Ogievetsky, and E.~Sokatchev,
\newblock Class. Quant. Grav. {\bf 2}, 617 (1985).

\bibitem{grisaru82}
M.~T. Grisaru and W.~Siegel,
\newblock Nucl. Phys. {\bf B201}, 292 (1982).

\bibitem{buchbinder97}
I.~L. Buchbinder, E.~I. Buchbinder, S.~M. Kuzenko, and B.~A. Ovrut,
\newblock Phys. Lett. {\bf B417}, 61 (1998).

\bibitem{buchbinder97:2}
I.~L. Buchbinder, S.~M. Kuzenko, and B.~A. Ovrut,
\newblock Phys. Lett. {\bf B433}, 335 (1998).

\bibitem{clark78}
T.~E. Clark, O.~Piguet, and K.~Sibold,
\newblock Nucl. Phys. {\bf B143}, 445 (1978).

\bibitem{sohnius79}
M.~F. Sohnius,
\newblock Phys. Lett. {\bf B81}, 8 (1979).

\bibitem{fisher83}
A.~W. Fisher,
\newblock Nucl. Phys. {\bf B229}, 142 (1983).

\bibitem{jones75}
D.~R.~T. Jones,
\newblock Nucl. Phys. {\bf B87}, 127 (1975).

\bibitem{howe83}
P.~S. Howe, K.~S. Stelle, and P.~C. West,
\newblock Phys. Lett. {\bf B124}, 55 (1983).

\bibitem{novikov83}
V.~A. Novikov, M.~A. Shifman, A.~I. Vainshtein, and V.~I. Zakharov,
\newblock Nucl. Phys. {\bf B229}, 381 (1983).

\bibitem{fisher84}
A.~W. Fisher,
\newblock Nucl. Phys. {\bf B241}, 243 (1984).

\bibitem{zumino79}
B.~Zumino,
\newblock Phys. Lett. {\bf B87}, 203 (1979).

\bibitem{alvarez81}
L.~Alvarez-Gaume and D.~Z. Freedman,
\newblock Commun. Math. Phys. {\bf 80}, 443 (1981).

\bibitem{hughes86}
J.~Hughes, J.~Liu, and J.~Polchinski,
\newblock Phys. Lett. {\bf B180}, 370 (1986).

\bibitem{hughes86:2}
J.~Hughes and J.~Polchinski,
\newblock Nucl. Phys. {\bf B278}, 147 (1986).

\bibitem{bagger94}
J.~Bagger and A.~Galperin,
\newblock Phys. Lett. {\bf B336}, 25 (1994).

\bibitem{bagger97}
J.~Bagger and A.~Galperin,
\newblock Phys. Lett. {\bf B412}, 296 (1997).

\bibitem{antoniadis96}
I.~Antoniadis, H.~Partouche, and T.~R. Taylor,
\newblock Phys. Lett. {\bf B372}, 83 (1996).

\bibitem{striebel87}
M.~Striebel,
\newblock Magnetic {M}onopoles in a {C}onstant {B}ackground {G}auge {F}ield,
\newblock PhD thesis at the {U}niversity of {B}ern, 1987.

\bibitem{minkowski90}
P.~Minkowski,
\newblock Czechoslovak Journal of Physics {\bf B40}, 1003 (1990).

\bibitem{minkowski90:2}
P.~Minkowski,
\newblock Phys. Lett. {\bf B237}, 531 (1990).

\bibitem{callan76}
J.~Curtis G.~Callan, R.~F. Dashen, and D.~J. Gross,
\newblock Phys. Lett. {\bf B63}, 334 (1976).

\bibitem{jackiw76}
R.~Jackiw and C.~Rebbi,
\newblock Phys. Rev. Lett. {\bf 37}, 172 (1976).

\bibitem{hooft76}
G.~'t~Hooft,
\newblock Phys. Rev. Lett. {\bf 37}, 8 (1976).

\bibitem{hooft76:2}
G.~'t~Hooft,
\newblock Phys. Rev. {\bf D14}, 3432 (1976).

\bibitem{rajaraman82}
R.~Rajaraman,
\newblock {\em Solitons and {I}nstantons},
\newblock North-Holland Publishing Company, 1982.

\bibitem{bernard77}
C.~W. Bernard and E.~J. Weinberg,
\newblock Phys. Rev. {\bf D15}, 3656 (1977).

\bibitem{rothe79}
H.~J. Rothe and J.~A. Swieca,
\newblock Nucl. Phys. {\bf B149}, 237 (1979).

\bibitem{witten79}
E.~Witten,
\newblock Nucl. Phys. {\bf B156}, 269 (1979).

\bibitem{veneziano79}
G.~Veneziano,
\newblock Nucl. Phys. {\bf B159}, 213 (1979).

\bibitem{shifman80}
M.~A. Shifman, A.~I. Vainshtein, and V.~I. Zakharov,
\newblock Nucl. Phys. {\bf B166}, 493 (1980).

\bibitem{banerjee94}
H.~Banerjee, P.~Mitra, and D.~Chatterjee,
\newblock Z. Phys. {\bf C62}, 511 (1994).

\end{thebibliography}
\end{document}